\documentclass[a4paper,12pt]{article} 

\usepackage[english]{babel} 

\usepackage{amssymb}
\usepackage{setspace}
\usepackage{afterpage}
\usepackage{latexsym}
\usepackage{amsmath}
\usepackage{amsthm}
\usepackage{amsfonts}
\usepackage{mathrsfs}
\usepackage{graphicx}
\usepackage{hyperref}

\newcommand{\lR}{\mathrm{I\hspace{-0.7mm}R}}

\numberwithin{equation}{section}

\hypersetup{ pdftitle={Draft},
pdfauthor={Bobby Eka Gunara and Ainol Yaqin}, 
pdfkeywords={Nonminimal Derivative Coupling Gravity}, 
bookmarksnumbered, pdfstartview={FitH}, urlcolor=blue, }

\begin{document}
\pagestyle{plain}


 \title{\LARGE \textbf{Comments On "Self-Gravitating Spherically Symmetric Solutions in Scalar-Torsion Theories" }}

\author{{ Ainol Yaqin$^{\sharp}$ and Bobby Eka Gunara$^{\flat,\sharp}\footnote{Corresponding author}$ } \\
$^{\flat}$\textit{\small Indonesian Center for Theoretical and
Mathematical Physics (ICTMP)}
\\ {\small and} \\
$^{\sharp}$\textit{\small Theoretical Physics Laboratory}\\
\textit{\small Theoretical High Energy Physics and Instrumentation Research Group,}\\
\textit{\small Faculty of Mathematics and Natural Sciences,}\\
\textit{\small Institut Teknologi Bandung}\\
\textit{\small Jl. Ganesha no. 10 Bandung, Indonesia, 40132}\\
\small email: yaqin.al@students.itb.ac.id, bobby@fi.itb.ac.id\\
}

\date{}

\maketitle

\begin{abstract}
 We find a crucial miscalculation in \cite{Kofinas:2015hla} which leads to the wrong master equation. This follows that there is no wormhole-like solution for hyperbolic scalar potential and the solution at large distances differs from that of \cite{Kofinas:2015hla}. 
 \end{abstract}

\section{Master Equation}
\label{sec:mastereq}
Let us start our discussion to section III in \cite{Kofinas:2015hla}. After redefining some new variables $(x, y, z)$, the field equations transform into 
\begin{eqnarray} 
&& \dot{y}+\frac{3}{2} y^{2} - \frac{e^{-2x}}{2\nu ^{2} } + \frac{\kappa ^{2}  \left(\varepsilon \nu ^{2} + 2V  \right)  }{4\nu ^{2} \left(1- 2\xi \kappa ^{2} \nu ^{2} \right)} = 0  ~ , \nonumber\\
yz &=& \frac{\kappa ^{2} }{\nu ^{2} \left(1-6\xi \nu ^{2} \kappa ^{2} \right)} \left( \frac{   (1- 2 \xi \nu ^{2} \kappa ^{2}) }{\kappa ^{2} } e^{-2x} + \frac{\nu ^{2}}{2}  - V  \right)  \equiv F ~ , \nonumber\\
&& \dot{z}+\frac{1}{2} z^{2} + \frac{e^{-2x}}{2\nu ^{2} }  +\frac{3 \kappa ^{2} \left( \nu ^{2} + 2V  \right) }{4\nu ^{2} \left(1-2\xi \kappa ^{2} \nu ^{2} \right)} = 0      ~ ,  \label{simplyEOM2}
\end{eqnarray} 
where $\varepsilon = \pm 1$. For the case at hand we choose $\varepsilon = + 1$ since we have a non-phantom scalar field \cite{Kofinas:2015hla}. Next, we introduce again a set of new variables
\begin{eqnarray}
 Y &\equiv& y^2  ~ , \nonumber \\ 
 Z &\equiv& z^2  ~ ,
 \end{eqnarray}
such that we could have 
\begin{eqnarray}
\frac{1}{2} \frac{dY}{dx}  &=& \dot{y} ~ , \nonumber \\ 
 \frac{1}{2} \varsigma \sqrt{ \frac{Y}{Z}} \frac{dZ}{dx}   &=&  \dot{z}  ~ , \label{thecorrectone}
 \end{eqnarray}
where $\varsigma = \pm 1$.  Then, (\ref{simplyEOM2}) can be rewritten as
\begin{eqnarray}
\frac{dY}{dx} + 3Y - \frac{e^{-2x}}{\nu^{2} } + \frac{\kappa ^{2} \left( \nu ^{2}  + 2V  \right) }{2 \nu ^{2} \left(1- 2\xi \kappa ^{2} \nu ^{2} \right)}  &=& 0 ~ , \nonumber \\ 
YZ  &=& F^2 ~ , \nonumber \\ 
\varsigma \sqrt{\frac{Y}{Z}} ~ \frac{dZ}{dx} + Z +  \frac{e^{-2x}}{\nu^{2} }  +\frac{3\kappa^{2}  \left( \nu ^{2}  + 2V  \right) }{2\nu ^{2} \left(1- 2\xi \kappa ^{2} \nu ^{2} \right)} &=& 0  ~ . \label{simplyEOM3}
\end{eqnarray}
The third equation in (\ref{simplyEOM3}) differs from that of eq. (A5) in \cite{Kofinas:2015hla}. The crucial miscalculation is the missing prefactor $\varsigma \sqrt{ \frac{Y}{Z}}$ in the second equation of (\ref{thecorrectone}). Thus, they obtained the wrong master equation as we will see below.\\
\indent After some computations using all equations in (\ref{simplyEOM3}) we get the correct master equation
\begin{eqnarray}
&& 2\varsigma \frac{d^2Y}{dx^2} + \frac{1}{Y} \left(\frac{\tilde{\eta}}{2\eta} - \varsigma \right)\left(  \frac{dY}{dx} \right)^2 +   \left(\frac{2\nu^2 (\tilde{\eta} - \varsigma \eta)}{Y} + \frac{3\tilde{\eta}}{\eta}  + 3 \varsigma - \frac{6\eta}{\tilde{\eta}}\right)  \frac{dY}{dx}  \nonumber\\
&& \quad + \left(\frac{9\tilde{\eta}}{2\eta} -  \frac{18\eta}{\tilde{\eta}} \right) Y  + 6  \tilde{\eta}\nu^2 + \frac{2\eta \tilde{\eta} \nu^4}{Y} + \frac{8\eta}{\tilde{\eta}\nu^2} e^{-2x}  = 0 ~ , \label{HNLODE}
\end{eqnarray}
where $\eta$ and $\tilde{\eta }$  are constant defined as
\begin{eqnarray}
\eta &\equiv&  \frac{\kappa ^{2} }{2\nu ^{2} \left(1 - 2\xi \nu ^{2} \kappa ^{2}  \right)}   ~  ,\nonumber\\
\tilde{\eta }  &\equiv& \frac{\kappa ^{2} }{\nu ^{2} \left(1 - 6 \xi \nu ^{2} \kappa ^{2} \right)} ~ .
\end{eqnarray}
We also have 
\begin{eqnarray} 
\left(\frac{dx}{d\phi } \right)^{2} &=& Y(x)  ~  ,\nonumber\\
\left[\frac{d\ln \left(R N^2\right)}{dx} \right]^{2}  &=& \frac{Z}{Y}   ~  , \label{derivatphi}
\end{eqnarray} 
such that the static metric (3.1) in \cite{Kofinas:2015hla} can be written down as
\begin{equation} \label{gengeom} 
ds^{2} = -N^2  dt^{2} +\frac{dR^{2} }{\nu ^{2} R^{2} Y} + R^{2}  d\Omega^{2}  ~ , 
\end{equation} 
where $d\Omega^{2}$ is the 2-sphere metric.
\section{Solutions}
\label{sec:solbond}

\subsection{Solutions At Large Distances}
 In the asymptotic limit, namely  $x \to + \infty$ the function $Y(x)$ can be written as $Y(x) = Y_0 + Y_1 + \dots$ such that $|Y_0|   \gg |Y_1(x)| $. Such a setup simplifies (\ref{HNLODE}) into a linearized differential equation. At the zeroth order, we have
\begin{equation} 
Y_0 =  \frac{2 \eta \nu^2 }{ 3 ( 2 \sigma  -1 )} ~ ,
\end{equation} 
where
\begin{equation} 
\sigma \equiv \frac{\varepsilon \eta}{\tilde{\eta}} ~ ,
\end{equation} 
with  $ \varepsilon = \pm 1$. \\
\indent The function $Y_1$ satisfies the following linear equation
 \begin{eqnarray} 
\frac{d^2Y_1}{dx^2} + \frac{3}{2}\varsigma\varepsilon \left(\frac{\tilde{1+ \varsigma\varepsilon \sigma - 2\sigma^2}}{\sigma}  \right) \frac{dY_1}{dx} + \frac{9}{4}\varsigma\varepsilon \left(\frac{1-4\sigma^2}{\sigma}   \right) Y_1  + \varsigma\varepsilon \frac{4\sigma}{\nu^2}   e^{-2x} = 0 ~  . \label{linearHNLODE}
 \end{eqnarray} 
The general solution of (\ref{linearHNLODE}) is given by 
 \begin{eqnarray} 
Y_1 &=& C_1 ~ e^{\lambda_1 x} +  D_1 ~ e^{\lambda_2 x} -  \frac{4 \varsigma\varepsilon \sigma  e^{-2x}}{\nu^2(\lambda_1 +2) (\lambda_2 +2)}  \nonumber\\
~ , \label{sollinearHNLODE}
 \end{eqnarray} 
where $C_1, D_1 \in \lR$ with
 \begin{equation} 
 \lambda_{1, 2} = - \frac{3\varsigma\varepsilon}{4\sigma} \left( 1 + \varsigma\varepsilon \sigma - 2\sigma^2 \pm \left[ \left(1 + \varsigma\varepsilon \sigma - 2\sigma^2 \right)^2 + 4\varsigma\varepsilon \sigma \left( 1- 4\sigma^2 \right)   \right]^{1/2}\right) ~  , \label{eigenwert}
 \end{equation} 
%
%
The values of $\lambda_1$ or $\lambda_2$ should be negative, since $\lim_{ x \to +\infty} Y_1(x) \to 0$. So, the solution (\ref{linearHNLODE}) belongs to the following two cases. First, for $\varsigma\varepsilon =1$ we have $0 < \sigma < \frac{1}{2}$ and then, $-\frac{3}{2} < \lambda_1 < 0 $ and $\lambda_2 < -2$. Second, in the case of  $\varsigma\varepsilon =-1$  we have  $-\frac{1}{2}< \sigma < 0$ which follows $\lambda_1 < -2$ and $-\frac{3}{2} < \lambda_2 < 0 $. 
 \indent Solving the second equation in (\ref{derivatphi}), we obtain the first order lapse function $f$ 
 \begin{equation} 
 f (R) =  A_0 Y_0^{\varsigma \varepsilon/2\sigma}  R^{3\varsigma \varepsilon - 1 } \left( 1+ \frac{\varsigma  \varepsilon Y_1}{\sigma Y_0}  \right)  ~ . \label{lapsefunct1}
 \end{equation} 
 while the first equation gives us  
 \begin{eqnarray} 
  \phi (R) &=&  \phi_0 + \int \frac{dx}{\sqrt{Y_0 + Y_1}} \nonumber\\
  &\approx&  \phi_0 + Y_0^{-1/2} ~ {\mathrm{ln}}R -\frac{1}{2} Y_0^{-3/2} \Bigg( \frac{C_1}{\lambda_1}  ~ R^{\lambda_1} +  \frac{D_1}{\lambda_2} ~ R^{\lambda_2}\nonumber\\
  &&  + \frac{ 2 \varsigma\varepsilon \sigma}{\nu^2 (\lambda_1 +2) (\lambda_2 +2)}  R^{-2}    \Bigg) ~ .
  \end{eqnarray} 
The first order scalar potential can be obtained from the first equation in (\ref{simplyEOM3}), namely
\begin{eqnarray} 
V &=&  -\frac{\nu^2 (2\sigma +7)}{2 (2\sigma -1)}  -\frac{2}{\eta } \left( C_1 (\lambda_1 +3) ~ R^{\lambda_1} +  D_1 (\lambda_2 +3) ~R^{\lambda_2}  \right) \nonumber\\
&& + \left(\frac{4 \varsigma\varepsilon \sigma }{ \nu^2 (\lambda_1 +2) (\lambda_2 +2)} - \frac{1}{\nu^2}  \right)   R^{-2} ~ .\nonumber\\
\end{eqnarray} 
Looking back at the lapse function (\ref{lapsefunct1}), its lowest order leads to spaces of constant scalar curvature whose Ricci scalar has the form
 \begin{equation} 
 R = - (9 + 3 \varsigma \varepsilon) \nu^2 Y_0 ~ .
 \end{equation} 
For $\varsigma\varepsilon =1$ we have asymptotically AdS spacetime, while in the case of $\varsigma\varepsilon =-1$ the asymptotic spacetime is not Einstein. 

%



\subsection{No Wormhole-like Solutions}
Now, let us consider a case with scalar potential
\begin{equation} 
V(x) = \alpha e^{-2x} + \beta ~ , \label{Vexp}
 \end{equation} 
where $\alpha$ and $\beta$ are real constants. As $x \to +\infty$, we should have
\begin{equation} 
 \beta =  -\frac{\nu^2 (2\sigma +7)}{2 (2\sigma -1)} ~ . \label{betafix}
 \end{equation}
Inserting (\ref{Vexp}) into the first equation in (\ref{simplyEOM3}) we then have
\begin{equation} 
 Y = c_1 e^{-3x} + \frac{e^{-2x}}{\nu^2} - \frac{\eta \nu^2}{2} -2\eta\alpha e^{-2x} - \frac{2}{3} \eta \beta ~ . \label{Ysol}
 \end{equation}
Finally, inserting (\ref{Ysol}) into the master equation (\ref{HNLODE}) we conclude that for $c_1 =0$ there is no solution for $\alpha$ real.

%


\section*{Acknowledgments}

The work in this paper is supported by Riset KK ITB 2017 and Riset Desentralisasi DIKTI-ITB 2017.




\end{document}